\documentclass[12pt]{article}
\usepackage{amssymb,amsmath}
\usepackage[dvips]{graphicx}
\usepackage[dvips]{color}
\usepackage[mathscr]{eucal}

\begin{document}

\title{
$~$
\vspace{-25mm}
\begin{flushright}
{\small
\hfill hep-th/0210169
\\}
\end{flushright}
\vspace{5mm}
\textbf{On spectral density of Neumann matrices}
}
\author{
Dmitri Belov$\,{}^{1}$\footnote{On leave from Steklov Mathematical
Institute, Moscow, Russia}
\enspace and  Anatoly Konechny$\,{}^{2}$
\vspace{0.6cm}
\\
${}^{1}$  Department of Physics, Rutgers University\\
136 Frelinghuysen Rd.\\
Piscataway, NJ 08854, USA\\
\texttt{belov@physics.rutgers.edu}
\\
\\
${}^{2}$  Racah Institute of Physics,
The Hebrew University\\
Jerusalem, 91904, Israel\\
\texttt{tolya@phys.huji.ac.il}
}
\date{~}

\maketitle
\thispagestyle{empty}

\begin{abstract}
In hep-th/0111281 the complete set of eigenvectors and eigenvalues of Neumann
matrices was
found. It was shown also that the spectral density contains a divergent constant
piece
that being regulated by  truncation at level $L$ equals $\frac{\log L}{2\pi}$.
In
this paper
we find an exact analytic expression for the finite part of the spectral
density.
This function allows one to calculate finite parts of various determinants
arising
in string field theory computations. We put our result to some consistency
checks.
\end{abstract}

\newpage
%\tableofcontents

%%%%%%%%%%%%%%%%%%%%%%%%%%%%%%%%%%%%%%%%%%%%%%%%%%%%%%%%
\section{Introduction}
\setcounter{equation}{0}

The basic ingredient in the construction of the covariant
string field theory is Witten's
star product \cite{Witten}.
In the algebraic formulation this star product is
specified by a string three-vertex. This vertex has factorized form and
its matter part has the following expression
\begin{multline}
|V_{3}\rangle^{\textrm{matter}} = \int d^{26}p^{(1)}d^{26}p^{(2)}d^{26}p^{(3)}
\delta(p^{(1)}+p^{(2)}+p^{(3)})\exp\left[-\frac{1}{2}\sum_{r,s=1}^{3}
\sum_{m,n=1}^{\infty}a_{n}^{(r) \dagger}V_{nm}^{\prime\, rs}a_{m}^{(s)\dagger}
\right.
\\
\left.
-\frac{1}{\sqrt{2}}\sum_{r,s=1}^{3}\sum_{n=1}^{\infty}
p^{(r)}V_{0n}^{\prime\, rs}a^{(s)\dagger}
-\frac{1}{4}V_{00}^{\prime} \sum_{r}(p^{(r)})^{2}
\right] \bigotimes_{r=1}^{3}|p^{(r)}\rangle \, .
\label{vert}
\end{multline}
The infinite matrices $V^{\prime rs}$ are called Neumann matrices.
Let $C$ be the twist matrix: $C_{mn}^{\prime} = (-1)^{n}\delta_{mn}$. Denote
$$
M^{\prime\, rs}_{nm} = (C^{\prime\, }V^{\prime\, rs})_{nm} \, , \enspace
n,m=1,2,\dots
$$
The matrices  $M^{rs}$ commute with each other, have real entries and are
symmetric.
Their spectrum was found in \cite{spectroscopy}. The set of its eigenvectors
$v_{n}^{(\kappa)}$ is labeled by a continuous parameter $-\infty< \kappa
<\infty$. The value of this parameter is an eigenvalue of the operator
\begin{equation*}
K_1=L_1+L_{-1}
\end{equation*}
which commutes with matrices $M^{\prime\,rs}$.
We have
$$
\sum_{n=1}^{\infty}M^{\prime\, rs}_{mn}v_{n}^{(\kappa)} =
\mu^{rs}(\kappa)v_{m}^{(\kappa)}
$$
where the eigenvalues $\mu^{rs}(\kappa)$ are
\begin{equation}
\mu^{rs}(\kappa)=\frac{1}{1+2\cosh\frac{\pi\kappa}{2}}
\Bigl[1-2\delta_{r,s}+e^{\frac{\pi\kappa}{2}}\delta_{r+1,s}
+e^{-\frac{\pi\kappa}{2}}\delta_{r,s+1}\Bigr].
\end{equation}
The eigenvectors $v_{n}^{(\kappa)}$ are given by their generating function
\begin{equation} \label{gen_fn}
f^{(\kappa)}(z)=\sum_{n=1}^{\infty}\frac{v_{n}^{(\kappa)}}{\sqrt{n}}z^{n} =
\frac{1}{\kappa\sqrt{{\cal N}(\kappa)}}(1-e^{-\kappa\tan^{-1}z})
\end{equation}
where
$$
{\cal N}(\kappa)=\frac{2}{\kappa}\sinh\left(\frac{\pi\kappa}{2}\right) \, .
$$

It was also shown in \cite{spectroscopy}, \cite{Okuyama1} that this set of
eigenvectors is orthogonal and complete
\begin{subequations}
\label{o_c}
\begin{align}
\sum_{n=1}^{\infty} v_{n}^{(\kappa)}v_{n}^{(\kappa')}&=\delta(\kappa -\kappa')
\,
,
\\
\int_{-\infty}^{+\infty}d\kappa\, v_{n}^{(\kappa)}v_{m}^{(\kappa)} &=
\delta_{n,m} \,
.
\end{align}
\end{subequations}

Let us also note that the spectral representation for Neumann coefficients
$V_{0n}^{\prime\,rs}$ can be also worked out using the standard relations
between these
coefficients and matrices $M^{\prime\,rs}$. In the basis $v_{n}^{(\kappa)}$
these
coefficients can be considered as vectors in $l_{\infty}$ space and the
coordinates
for these vectors with respect to the basis $v_{n}^{(\kappa)}$ can be found
(see for example \cite{BK1} formulae (3.9), (3.10)). Notice that
the vertex \eqref{vert} can also be written in the alternative form, which
can be obtained by integrating over momentum.
In this case the vertex is specified by matrices $M^{rs}$
whose matrix indices run from $0$ to $\infty$. The spectral representation
of these matrices was found in \cite{0202176,Dima1}.
In the present  paper we will construct spectral density
related to matrices $M^{\prime\,rs}$.

The spectral measure for spectral parameter $\kappa$ was first considered in
\cite{spectroscopy}. It was shown that if one truncates the matrix
$(K_{1})_{mn}$
to a finite $L\times L$ matrix by restricting $n,m=1, \dots L$ the eigenvalues
have a uniform distribution with density
$$
\rho^{L}(\kappa) \sim \frac{\log L}{2\pi}
$$
in the limit $L\to \infty$. There are however corrections to the density
at finite $L$ and as it was already noted in \cite{spectroscopy} these
corrections
become large for large $|\kappa|$ (and fixed $L$).

In \cite{BK1} we presented a numerical study of the spectral density function
that
indicates that there is a nontrivial part $\rho_{\textsf{fin}}(\kappa)$ of the
spectral
density that stays finite in the limit $L\to \infty$.
More precisely using the orthogonality and completeness relations (\ref{o_c})
one can write the following expression for  the level truncated distribution
function
\begin{equation}
\rho^{L}(\kappa) = \sum_{n=1}^{L} v_{n}^{(\kappa)}v_{n}^{(\kappa)} \, .
\end{equation}
This expression stems from the fact that for a given symmetric matrix $A_{nm}$
with
spectral
representation $A(\kappa)$ the following identity is true
\begin{equation} \label{rel}
\sum_{n=1}^{L}A_{nn} = \int_{-\infty}^{\infty}\rho^{L}(\kappa) d\kappa\,
A(\kappa)
\, .
\end{equation}
One can then numerically study  the difference $\rho^{L}(\kappa) - \frac{\log
L}{2\pi}$.
The numerical evidence we obtained in \cite{BK1} suggests that for large
$L$
this difference converges, at least in the vicinity of $\kappa=0$ to a well-
defined
continuous function. We will formally define this function as
\begin{equation} \label{rho_formal}
\rho_{\textsf{fin}}(\kappa) =
\lim_{L\to \infty} \rho^{2L}(\kappa) - \frac{1}{2\pi}\sum_{n=1}^{L}\frac{1}{n}
\end{equation}
where the limit $2L$ instead of $L$ and the use of the sum of $1/n$ instead of
the logarithm
are dictated solely by technical convenience.

By using the known formula for coefficients $v_{n}^{(0)}$
it is easy to show analytically that $\rho_{\textsf{fin}}(0)=\frac{\log
2}{\pi}$.

Our goal in this paper is to obtain an exact analytic expression for
$\rho_{\textsf{fin}}(\kappa)$. To achieve that goal we use a separate
regularization
for both terms in (\ref{rho_formal})  defining
\begin{equation} \label{rho_q}
\rho^{q}(\kappa) = \sum_{n=1}^{\infty} v_{n}^{(\kappa)}
q^{n}v_{n}^{(\kappa)}
\end{equation}
and
$$
\rho_{\textsf{fin}}^{q}(\kappa) = \rho^{q}(\kappa) -
\frac{1}{2\pi}\sum_{n=1}^{\infty}
\frac{q^{2n}}{n} = \rho^{q}(\kappa) + \frac{1}{2\pi}\log(1-q^{2})
$$
where $q$ is a positive number less than one. We are looking then for a limit
$q\to 1$ of this expression.
In fact it is technically more advantageous, but leads to the same result,
if we consider the limit $q\to 1$ of the quantity
\begin{equation}\label{rho_finq}
\rho_{\textsf{fin}}^{q} = \rho^{q}(\kappa) - \rho^{q}(0) + \frac{\log 2}{\pi} \,
.
\end{equation}
The computation of this quantity is done in the next section by performing
a certain  contour integration. Taking the limit $q\to 1$ one obtains
\begin{equation} \label{main1}
\rho_{\textsf{fin}}(\kappa) = \frac{2\log2-\gamma_E}{2\pi} -
\frac{1}{4\pi}\left[ \psi\Bigl(1 + \frac{\kappa}{2i}\Bigr)
+ \psi\Bigl(1 -\frac{\kappa}{2i}\Bigr)\right]
\end{equation}
where $\gamma_E$ is the Euler constant and $\psi(z)$ is the logarithmic
derivative
of the $\Gamma$-function. Formula (\ref{main1}) arises in computations as
an integral representation
\begin{equation} \label{main2}
\rho_{\textsf{fin}}(\kappa) = \frac{\log 2}{\pi} +
\frac{1}{2\pi}\int_{0}^{\infty}dt\,
\frac{\cos\left(\frac{\kappa t}{2}\right) -1}{e^{t}-1} \, .
\end{equation}

Formulae (\ref{main1}), (\ref{main2}) are the main result of this paper.
A particular   SFT computation involving $\rho_{\textsf{fin}}(\kappa)$ is a computation
of
an overlap of two surface states discussed in \cite{BK1}. Even with the use
of the exact analytic expression (\ref{main1}) this computation still seems
to be at odds with CFT results \cite{peskin2}. We hope to return to this issue
in the nearest future.

%%%%%%%%%%%%%%%%%%%%%%%%%%%%%%%%%%%%%%%
\section{Computation}
\setcounter{equation}{0}

\begin{figure}[t]
\centering
\includegraphics[width=400pt]{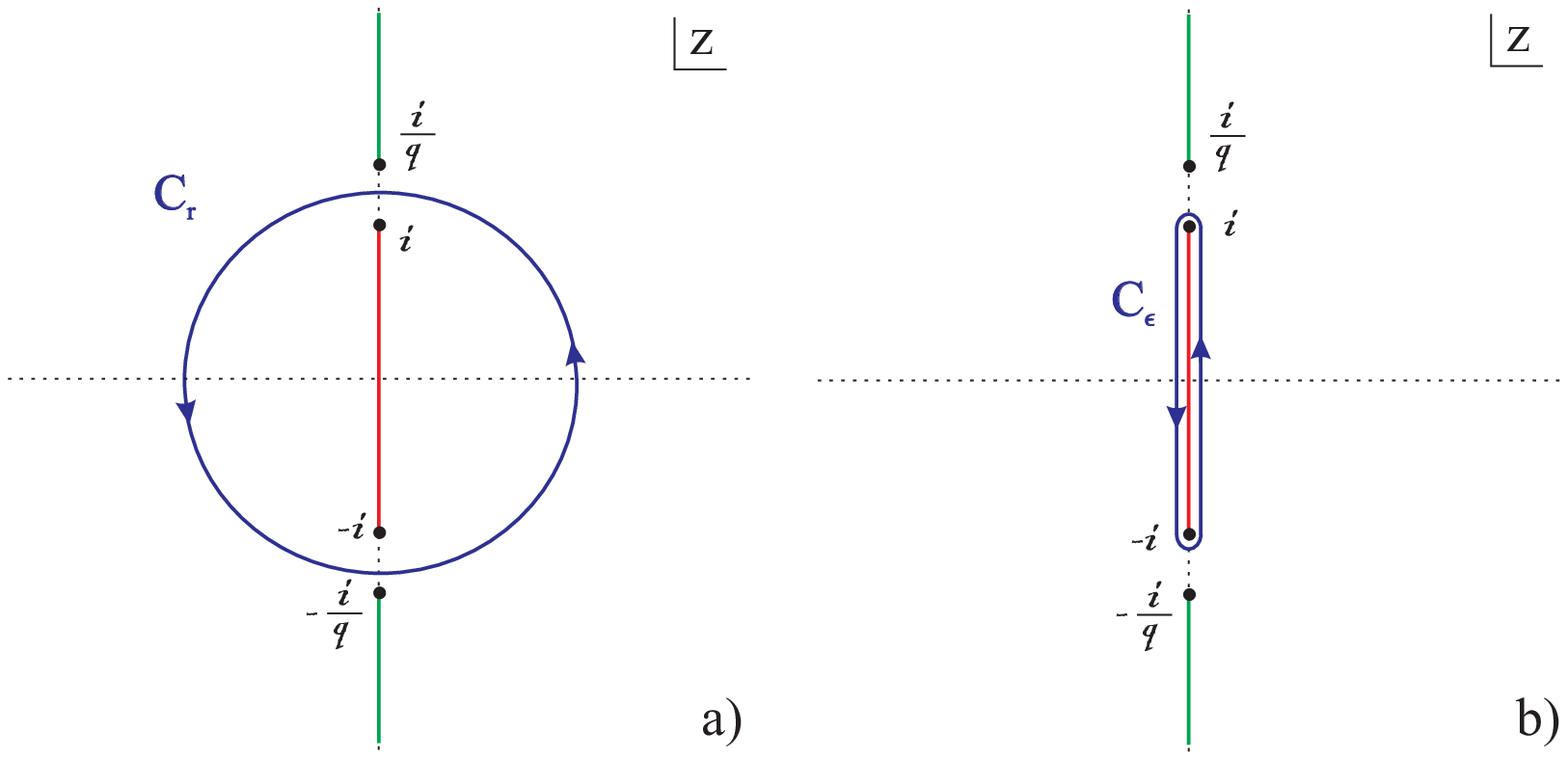}
\caption{}
\label{fig:1}
\end{figure}

We need to find an analytic expression for the regulated spectral
density $\rho^{q}(\kappa)$ defined in (\ref{rho_q}). Let us start by computing
a more general expression
$$
\rho^{q}(\kappa, \kappa') =
\sum_{n=1}^{\infty}v_{n}^{(\kappa)}q^{n}v_{n}^{(\kappa')} \, .
$$
It can be expressed via a contour integral (see Figure~\ref{fig:1}a)
\begin{equation}\label{rho_kk}
\rho^{q}(\kappa,\kappa')=\frac{q}{2\pi i}\oint_{C_{r}}
(\partial f)^{(\kappa)}(qz) f^{(\kappa')}\left( \frac{1}{z}\right)\, .
\end{equation}
Here
$$
(\partial f)^{(\kappa)}(qz) = \frac{1}{\sqrt{{\cal N}(\kappa)}}
\frac{1}{1 + (qz)^{2}}\exp( -\kappa \tan^{-1}(qz))
$$
is an analytic  function defined on a complex plane with  cuts going along the
imaginary axis from $i/q$ to $+i\infty$ and from $-i/q$ to $-i\infty$ and the
branch of
$$
\tan^{-1}(qz) = \frac{1}{2i}\log \left(\frac{1+iqz}{1-iqz} \right)
$$
is chosen by the standard series expansion within the circle $|z|<1/q$, i.e.
one takes main branches for both logarithms.
The function
$$
f^{(\kappa')}\left( \frac{1}{z}\right) =\frac{1}{\kappa'\sqrt{{\cal
N}(\kappa')}}
\left(1-\exp(-\kappa'\tan^{-1}(1/z))\right)
$$
is defined on a complex plane with a cut going along the imaginary axis from
 $-i$ to $i$ and the branch is fixed by the standard series expansion outside
 of the circle $|z|=1$. The contour $C_{r}$ in (\ref{rho_kk}) is chosen
 to be a circle centered at the origin and of radius $1<r<1/q$, as both
functions
 are holomorphic within the annulus $1<|z|<1/q$. The contour and the cuts are
depicted
 on Figure~\ref{fig:1}a. We can deform the circle $C_{r}$ to the contour
 $C_{\epsilon}=\{z=ix - \epsilon\}\cup\{z=ix+\epsilon\}
 \cup C_i\cup C_{-i}$, $ -1\le x\le 1$
depicted on Figure~\ref{fig:1}b. We obtain then
\begin{multline} \label{rho_kk2}
\rho^{q}(\kappa, \kappa') = \frac{q}{2\pi}\int_{-1}^{1} dx \,
(\partial f)^{(\kappa)}(qix) \left[
f^{(\kappa')}\left( \frac{1}{ix +\epsilon}\right) -
f^{(\kappa')}\left( \frac{1}{ix -\epsilon}\right)\right]
\\
=\frac{q}{\pi } \sqrt{ \frac{ {\cal N}(\kappa')}{{\cal N}(\kappa)}}
\int_{-1}^{1}\frac{dx}{1-(qx)^{2}}\exp\left[
\frac{i\kappa}{2}\log\left(\frac{1-qx}{1+qx}\right) +
\frac{i\kappa'}{2}\log\left(\frac{1+x}{1-x}\right)\right]
\end{multline}
where the last expression is evaluated  in the limit $\epsilon\to 0$, we used
the fact
 the integral over half circle of radius $\epsilon$ $C_{i}$ and $C_{-i}$
is zero and
$$
\tan^{-1}\left(\frac{1}{ix\pm 0}\right)= \frac{1}{2i}\log\left(
\frac{1+x}{1-x} \right) \pm \frac{\pi}{2} \, , \enspace -1\le x\le 1 \, .
$$
If at this point we make in (\ref{rho_kk2}) a substitution
$$
y=\frac{1}{2}\log\left( \frac{1-qx}{1+qx}\right)
$$
we arrive at the  expression
\begin{subequations}
\begin{equation}
\rho^{q}(\kappa, \kappa') = \frac{1}{2\pi}\sqrt{ \frac{ {\cal N}(\kappa')}{{\cal
N}(\kappa)}}
\int_{-R}^{R}dy\, \exp\bigl[iy(\kappa-\kappa')\bigr]
\exp\left[\frac{i\kappa'}{2} g_{q}(y) \right]
\end{equation}
where
\begin{equation}
g_{q}(y) =\log\left( \frac{(q-\tanh y)}{(q+\tanh y)} \frac{(1+\tanh y)}{(1-\tanh
y)}\right)
\end{equation}
and
\begin{equation}
R=R(q)={\rm arctanh}(q) \to \infty \, \enspace \mbox{as} \enspace q\to 1 \, .
\end{equation}
\end{subequations}
Using the last expression it can be shown  that
$$
\lim_{q\to 1}\rho^{q}(\kappa, \kappa')=\delta(\kappa-\kappa') \, .
$$
On a formal level
we note that the support of the function
$g_{q}(y)$
is pushed to $|y|=\infty$ as $q\to 1$. On the other hand if we consider an
integral
$\int_{-\infty}^{\infty} d\kappa \rho^{q}(\kappa, \kappa') t(\kappa) $
where $t(\kappa)$ is a Schwartz class test function, then
the Fourier transform $T(y)$ of the function $t(\kappa)/\sqrt{{\cal N}(\kappa)}$
provides a damping factor at $y\to \infty$ that allows one to drop out the
function $g_{q}(y)$ in the limit $q\to 1$.
This provides an alternative (and in our opinion a cleaner) derivation of
the orthogonality property (\ref{o_c}).

 This being noted we set now
 $\kappa=\kappa'$ in (\ref{rho_kk2}) and rewrite it
as
\begin{equation}
\rho^{q}(\kappa) = \frac{1}{\pi}\int_{0}^{q}\frac{dx}{1-x^{2}}
\cos\left[\frac{i\kappa}{2}\log
\left(\frac{1- \frac{1-q}{1+x}}{1 - \frac{1-q}{1-x}}\right)
\right]
\end{equation}
As the function
$$
\log\left(1-\frac{1-q}{1+x}\right)
$$
uniformly converges to zero on ${\mathbb R}_{+}$ in the limit $q\to 1$ we
can safely drop it in the above expression. After a substitution
$$
t=\log\left(1- \frac{1-q}{1-x} \right)
$$
we obtain using (\ref{rho_finq})
$$
\rho_{\textsf{fin}}(\kappa) = \frac{\log 2}{\pi} + \frac{1}{2\pi}
\lim_{q\to 1}\int_{\log q^{-1}}^{\infty}dt\, \frac{\cos\left(\frac{\kappa
t}{2}\right) -1}
{e^{t}-1 + \frac{q-1}{2}e^{t}}
$$
that uniformly converges to the function (\ref{main2}). This completes the
derivation of that formula.

%%%%%%%%%%%%%%%%%%%%%%%%%%%%%%%%%
\section{Verification of the analytic expression}
\setcounter{equation}{0}

In this section  we present two consistency checks  verifying that (\ref{main1})
gives  the right finite part of the spectral density.

\subsection{Numerical check}
\begin{figure}[t]
\centering
\includegraphics[width=370pt]{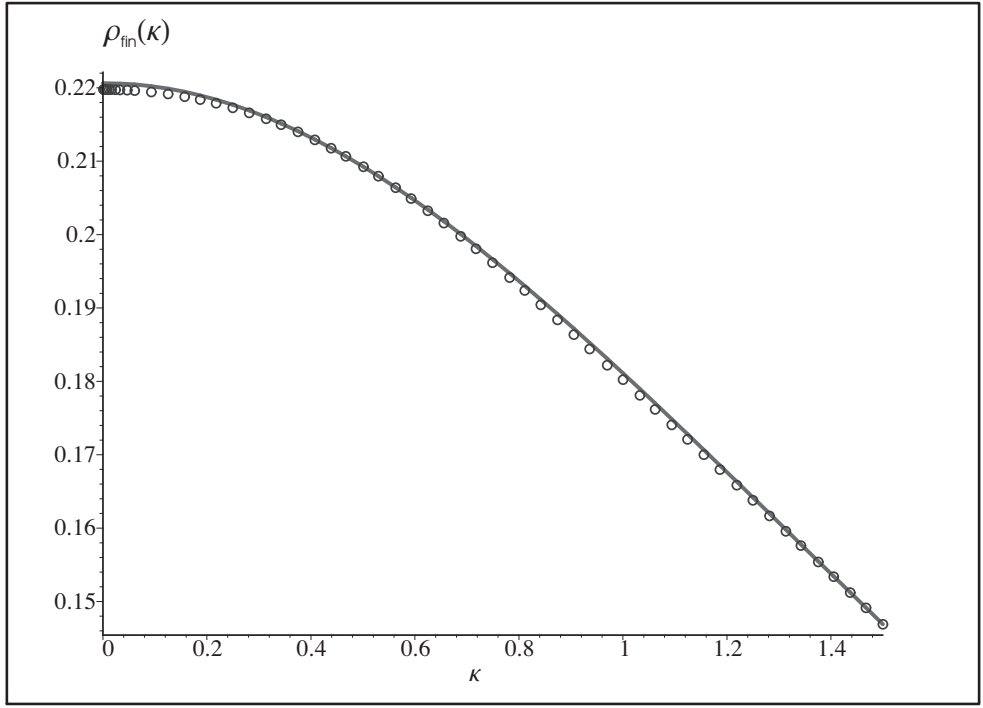}
\caption{The solid line represents a plot of analytic expression
for the finite part of the spectral density \eqref{main2}.
The circles represents
numerical results for the finite density obtained by evaluating
\eqref{rho_formal} for $L=91$.
}
\label{fig:2}
\end{figure}

Using equation \eqref{rho_formal} for finite value of $L$
one can calculate the finite part of the spectral density
in the vicinity of zero.
On Figure~\ref{fig:2} we present the result of
this calculation for $L=91$ (circled line). One
sees that the numeric result is in a very good
agreement with our analytic expression for the
finite part of the spectral density (solid line).
 The exact analytic expression also gives the asymptotic behavior of
 $\rho_{\textsf{fin}}(\kappa)$ not captured by the numerical results.
Namely  $\rho_{\textsf{fin}}\sim -\frac{1}{\pi}\log|\kappa| $
as $|\kappa| \to \infty$.

\subsection{Analytic check. Trace of operator $B$.}
In \cite{spectroscopy} a certain operator denoted by $B$ was introduced for
which
both the matrix elements and the eigenvalues are of a simple form
and known explicitly. Namely we have
\begin{equation*}
B_{nm} =
\left\{%
\begin{array}{ll}
    -\frac{(-1)^{\frac{m-n}{2}}\sqrt{nm}}{(m+n)^{2}-1}, &
    n+m\enspace\hbox{even;} \\
    0, & n+m\enspace \hbox{odd.} \\
\end{array}%
\right.
\end{equation*}
while the eigenvalues corresponding to eigenvectors $v^{(\kappa)}_{n}$ reads
$$
\beta(\kappa)= -\frac{1}{4}\frac{\pi \kappa}{\sinh\left(\frac{\pi
\kappa}{2}\right)} =
-\frac{\pi}{2{\cal N}(\kappa)} \, .
$$
We can use these representations to check the asymptotic form of relation
(\ref{rel}).
Namely from (\ref{rel}) we have
$$
\sum_{n=1}^{2L}B_{nn} = -\sum_{n=1}^{2L}\frac{n}{4n^{2}-1}=
 \int_{-\infty}^{+\infty}d\kappa\, \beta(\kappa)\left[
 \frac{1}{2\pi}\sum_{n=1}^{L}\frac{1}{n} + \rho_{\textsf{fin}}^{2L}(\kappa)
 \right]
$$
and therefore
\begin{equation} \label{1}
\int_{-\infty}^{+\infty}d\kappa \, \rho_{\textsf{fin}}(\kappa)\beta(\kappa) =
\lim_{L\to
\infty}
\left[\sum_{n=1}^{L}\frac{1}{4n} - \sum_{n=1}^{2L}\frac{n}{4n^{2}-1}
\right] =
\frac{1}{4}-\frac{3\log 2}{4} \, .
\end{equation}
On the other hand if  we use  our analytic expression (\ref{main2})
for $\rho_{\textsf{fin}}(\kappa)$ we have
\begin{equation*}
\int_{-\infty}^{+\infty}d\kappa \, \rho_{\textsf{fin}}(\kappa)\beta(\kappa) = \\
\frac{\log 2}{\pi}\int_{-\infty}^{+\infty}d\kappa\, \beta(\kappa) +
\frac{1}{2\pi}\int_{-\infty}^{+\infty}d\kappa\int_{0}^{\infty}dt\,
\frac{\beta(\kappa)(\cos\left(\frac{\kappa t}{2}\right)-1)}{e^{t}-1} \, .
\end{equation*}
The first integral can be evaluated in a straightforward way with the result
$-\frac{1}{2}\log 2$ while in the second term we can first evaluate the integral
over
$\kappa$ by summing up residues in the upper half plane:
$$
\int_{-\infty}^{+\infty}d\kappa \, \frac{e^{i\kappa t}}{{\cal N}(\kappa)} =
\frac{1}{\cosh^{2} t}
$$
and then take the integral over $t$
$$
\frac{1}{4}\int_{0}^{\infty}dt\, \frac{1-e^{-t}}{e^{t} + e^{-t} + 2} =
\frac{1}{4}-\frac{\log 2}{4} \, .
$$
Combining both terms together we obtain the same number as in (\ref{1}).

%%%%%%%%%%%%%%%%%%%%%%%%%%%%%%%%%%%%%%%%%%%%%%%%%%%

\vspace{1cm}
\textbf{Note added:}
While this paper was nearing completion, the paper \cite{0210155} appeared,
which contains in Section~3 the same  result  ~\eqref{main1} we obtained.
The calculations in that paper  are technically different from the ones we
present.

%%%%%%%%%%%%%%%%%%%%%%%%%%%%%%%%%%%%%%%%%%%%%%%%%%%%%%%%%%%
\section*{Acknowledgments}

The work of D.B. was supported in part by RFBR grant 02-01-00695.
The work of A.K. was supported in part by BSF-American-Israel Bi-National
Science
Foundation, the Israel Academy of Sciences and Humanities-Centers of Excellence
Program, the German-Israel Bi-National Science Foundation.

\end{document}